\documentclass[american,aps,pra,reprint, superscriptaddress]{revtex4-1}
\usepackage{amsmath,amscd,amsthm,amssymb}
\usepackage{mathrsfs}
\usepackage{graphicx}
\usepackage{epsf,epstopdf}
\usepackage{caption3}
\usepackage[all]{xy}
\usepackage{tikz}
\usepackage{enumerate}
\usepackage{mathtools}
\usepackage[unicode=true,pdfusetitle, bookmarks=true,bookmarksnumbered=false,bookmarksopen=false, breaklinks=false,pdfborder={0 0 0},backref=false,colorlinks=false] {hyperref}
\hypersetup{colorlinks,linkcolor=myurlcolor,citecolor=myurlcolor,urlcolor=myurlcolor}
\usepackage{graphics, times, braket, color, bm}
\usepackage[up]{subfigure}
\usepackage{cleveref}

\definecolor{myurlcolor}{rgb}{0,0,0.7}

\def\be{\begin{equation}}
\def\ee{\end{equation}}
\def\bea{\begin{eqnarray*}}
\def\eea{\end{eqnarray*}}
\def\ot{\otimes}

\theoremstyle{plain}

\providecommand{\theoremname}{Theorem}

\newcommand{\iinner}[2]{\langle #1 | #2\rangle}
\newcommand{\out}[2]{| #1\rangle\langle #2 |}
\DeclareMathOperator{\trace}{tr}
\newcommand{\ptr}[2]{\trace_{#1}({#2})}
\newcommand{\tr}[1]{\ptr{}{#1}}
\newcommand{\id}{\mathbb{I}}

\newcommand*{\myproofname}{Proof}

\makeatother
\newcommand{\DOI}[2]{\href{https://doi.org/#1}{#2}}

\def\cD{\mathcal{D}}
\def\cF{\mathcal{F}}\def\cH{\mathcal{H}}\def\cI{\mathcal{I}}

\def\rE{\mathrm{E}}

\theoremstyle{definition}

\theoremstyle{remark}

\graphicspath{{figs/}}

\begin{document}

\title{Partial coherence versus entanglement}

 \author{Sunho Kim}
 \email{Corresponding author: kimsunho81@hrbeu.edu.cn}
 \affiliation{School of Mathematical Sciences, Harbin Engineering University, Harbin 150001, People's Republic of China}

  \author{Chunhe Xiong}
\email{Corresponding author: xiongch@zucc.edu.cn}
 \affiliation{School of Computer and Computing Science, Hangzhou City University, Hangzhou 310015, People's Republic of China}
\affiliation{Interdisciplinary Center for Quantum Information, Department of Physics, Zhejiang University, Hangzhou 310027, People's Republic of China}

 \author{Shunlong Luo}
\email{Corresponding author: luosl@amt.ac.cn}
 \affiliation{Academy of Mathematics and Systems Science, Chinese Academy of Sciences, Beijing 100190, People's Republic of China}
\affiliation{School of Mathematical Sciences, University of Chinese Academy of Sciences, Beijing 100049, People's Republic of China}

\author{Asutosh Kumar}
 \email{Corresponding author: asutoshk.phys@gmail.com}
 \affiliation{P. G. Department of Physics, Gaya College, Magadh University, Rampur, Gaya 823001, India}
 \affiliation{Vaidic and Modern Physics Research Centre, Bhagal Bhim, Bhinmal, Jalore 343029, India}

\author{Junde Wu}
\email{Corresponding author: wjd@zju.edu.cn}
\affiliation{School of Mathematical Sciences, Zhejiang University, Hangzhou 310027, People's Republic of China}

\begin{abstract}

We study partial coherence and its connections with entanglement. First, we provide a sufficient and necessary condition for bipartite pure state transformation under partial incoherent operations: A bipartite pure state can be transformed to another one if and only if a majorization relationship holds between their partial coherence vectors. As a consequence, we introduce the concept of maximal partial coherent states in the sense that they can be used to construct any bipartite state of the same system via partial incoherent operations. Second, we provide a strategy to construct measures of partial coherence  by use of symmetric concave functions. Third, we establish some relationships between partial coherence and entanglement. We show that the minimal partial coherence under local unitary operations is a measure of entanglement for bipartite pure states, which can be extended to all mixed states by convex-roof. 
We also show that partial coherence measures are induced through maximal entanglement under partial incoherent operations for bipartite pure states.
There is a one-to-one correspondence between entanglement and partial coherence measures. 

\end{abstract}
\maketitle

\section{Introduction}

Entanglement is one of the most important quantum correlations \cite{EPR1935,Horodecki2009} which serves as a basic ingredient for many useful quantum protocols, such as quantum cryptography \cite{Ekert1991}, quantum superdense coding \cite{Bennett1992}, and quantum teleportation \cite{Bennett1993}. Various  notions of quantum correlations related to entanglement, such as Bell non-locality \cite{Bell,Brunner}, quantum discord \cite{Henderson2001,Ollivier2001,LuoD,LuoD1,D3}, and quantum steering \cite{QS1,QS2}, are also very prominent and useful in quantum information theory.
In particular, entanglement can be related to quantum discord via state extensions \cite{LuoD1,D3}.

Coherence is a fundamental feature of quantum mechanics and only in recent years a resource theory of coherence  has been developed \cite{Streltsov1,Streltsov2,Baumgratz}. Thereafter, a lot of work has been done to enrich this theory, including quantification of coherence \cite{Girolami,Yuan,Shao,Pires,Rana,Rastegin, Napoli2016,Winter2016,Yu,Luo,Xiong2019,Bu,Xiong2018,SKim1}, reciprocity and distribution of coherence \cite{Asutosh}, states with the maximal coherence \cite{Peng,Yao,Zhao}, role of coherence in quantum algorithms \cite{Hillery2016,Matera2016,Ahnefeld2022,SKim2}, wave-particle duality models \cite{Bagan,Bera,LuoWP1,LuoWP2}, and partial coherence  \cite{Luo}. See also Ref. \cite{Streltsov1} for an overview. We will study  partial coherence in bipartite systems and connect it with entanglement.

Both entanglement and coherence are derived from the superposition principle of quantum mechanics, and coherence theory is directly inspired by entanglement theory. Therefore, as expected, many results of coherence theory are strikingly similar to those of entanglement theory, including coherence measures \cite{Baumgratz,Winter2016,Napoli2016} and state transformations \cite{Du,Zhu}.
Though coherence and entanglement are operationally equivalent \cite{Streltsov3}, they are conceptually different. That is, they are not the same thing exactly.
In addition, the relationships  between coherence measures and other quantumness  measures are studied in Refs. \cite{Streltsov3,Xi,Hu,Killoran,SKim3,Ma,Streltsov4,Tan,Kim}. For instance, Aubrun {\sl et al}. recently showed that entanglement and superposition are equivalent concepts in any physical theory \cite{Aubrun2022}.

Unlike entanglement, coherence is well-defined in single quantum systems.
As a kind of coherence in bipartite quantum systems, partial coherence may be employed to investigate relationships between coherence and entanglement. This paper is devoted to this issue.

First, we introduce the concept of {\it partial coherence vector} and provide a sufficient and necessary condition for state conversion under partial incoherent operations (PIO): A bipartite pure state can be transformed to another one with PIO if and only if the partial coherence vector of the former is majorized by the latter. This extends a result for coherence transformation under incoherent operations \cite{Du,Zhu}, and is reminiscent of the entanglement transformation under local operations and classical communication (LOCC) \cite{Nielsen}.

Second, we establish a  correspondence between measures of partial coherence  of pure states and symmetric concave functions by showing that each symmetric concave function induces a partial coherence measure on pure states, and  conversely, the restriction of any partial coherence measure to pure states can be also derived from a symmetric concave function. This correspondence can be directly extended to all mixed states via convex-roof extension.

Third, we show that each partial coherence measure can induce an entanglement measure, which means that entanglement can be regarded as the minimal partial coherence under all local unitary operations. On the contrary, by means of the symmetric concave function corresponding to each entanglement measure, we can define the corresponding partial coherence measure with convex-roof extension. In this way, we establish a one-to-one correspondence between entanglement measures and measures of partial coherence.

This paper is structured as follows. In Sec. II, we recall various prerequisite  concepts, including partial coherence, entanglement, symmetric convex functions, and majorization, which will be used subsequently. In Sec. III, we establish a relationship between measures of partial coherence and symmetric concave functions.  In Sec. IV, we establish a one-to-one correspondence between  measures of partial coherence and entanglement measures by means of symmetric concave functions. In Sec. V, we summarize the results. Appendix provides the  details of proofs.

\section{Preliminaries}

 \subsection{Resource theory of coherence and entanglement}


Let us first review the resource theory of coherence \cite{Baumgratz}, partial coherence \cite{Luo}, and entanglement \cite{Horodecki2009}. Any resource theory is composed of ``free states" and ``free operations". Each free state cannot contain any resource and each free operation preserves the set of free states, that is, it maps free states to free states.

{\em Coherence}--Let $\cH$ be a finite dimensional Hilbert space with a fixed  orthonormal basis $\{\ket{i}\},$ which will serve as the reference basis (computational basis) for considering coherence. Free states of coherence theory are called incoherent states, whose density operators are diagonal in the computational basis. Thus
\begin{align*}
\cI=\Big \{\sum_ip_i\ket{i}\bra{i}: p_i\ge 0,\sum_ip_i=1\Big \}
\end{align*}
is the set of incoherent states (with respect to the basis $\{\ket{i}\}$). Free operations $\Phi$ of coherence theory are called incoherent operations which are completely-positive and trace-preserving (CPTP) maps admitting an incoherent Kraus representation. That is, $\Phi (\rho )=\sum _i K_i\rho K^{\dagger}_i$
with each Kraus operator $K_i$ satisfying
\begin{align*}
\frac{K_i\rho  K^{\dagger}_i}{\tr{K_i\rho K^{\dagger}_i}}\in\cI
\end{align*}
for every incoherent state $\rho \in {\cal I}.$

{\em Partial coherence}--Consider a bipartite quantum system $ab$  described by the Hilbert space $\cH_{ab}=\cH_a\otimes\cH_b$ with $d_a=\dim {\cH_a}$ and $d_b=\dim {\cH_b}.$ For a fixed basis $\{\ket{i}: i=1,2,\cdots, d_a\}$ on party $a$, the set of partial incoherent states (i.e., the free states of partial coherence theory) are defined as
\begin{align*}
\cI_a=\Big \{\sum_ip_i\ket{i}\bra{i}\otimes\rho_i: p_i\ge 0,\sum_ip_i=1, \rho _i \in \cD(\cH_b)\Big \},
\end{align*}
where $\cD(\cH_b)$ denotes the set of all states (pure or mixed) of party $b$. Any quantum operation on party $a$ with each Kraus operator preserving partial incoherent states is called a partial incoherent operation (PIO), which serves as a free operation of partial coherence theory.

The partial decohering operation acting on a bipartite state $\rho$ of the composite system $ab$ is defined as
$$\Delta(\rho)=\sum_i\ket{i}\bra{i}\otimes \bra{i}\rho\ket{i},$$
 which maps any state to a partial incoherent state. Here $\bra{i}\rho\ket{i}
 ={\rm tr}_a \rho (|i\rangle \langle i| \otimes {\bf 1}_b)$ is a state on party $b$. Obviously, a partial decohering operation is partial incoherent. In particular,  if $\dim \mathcal{H}_b=1$, then the partial coherence theory reduces to the standard coherence theory.

{\em Entanglement}--Entanglement theory provides a paradigm for resource theory \cite{Horodecki2009}, whose free states are separable states of the forms
$$\rho=\sum_ip_i\rho^i_a\otimes\rho^i_b,$$
where $p_i\ge0,\sum_ip_i=1,$  $\rho^i_a$ and $\rho^i_b$ are local states for parties $a$ and $b$, respectively. Denote the set of separable states by $\mathcal{S}.$
 Any bipartite state that cannot be written in the above separable form is called entangled. Local operations and classical communication (LOCC) cannot map a separable state to an entangled one, therefore, the set of LOCC operations are regarded as the free operations of entanglement theory.

{\em Resource theory}--In general, measures for resource in a resource theory are functions $R $ defined on state space and satisfy the following conditions:

(R1) {\sl Faithfulness}: $R (\rho)\geq 0$, and $R (\rho)=0$ if and only if $\rho \in \cF$, where $\cF$ is the set of all free states.

(R2a) {\sl Monotonicity under any free operation}: $R (\Sigma(\rho)) \leq R (\rho)$ for any free operation $\Sigma$; or

 (R2b) {\sl Monotonicity under selective measurement} $\{K_n\}$: $\sum_np_nR (\sigma_n)\leq R (\rho)$, where $\sigma_n = K_n\rho K_n^\dagger/p_n$ with $p_n = \tr{K_n\rho K_n^\dagger}$.

(R3) {\sl Convexity}: $R (\sum_iq_i\rho_i) \leq \sum_iq_iR (\rho_i)$ for $q_i\geq 0, \sum_iq_i=1$ and quantum states $\rho_i.$

We remark that  (R2b) together with (R3) imply (R2a). When (R2b) and (R3) are satisfied, it is called a monotone, while when (R1)-(R3) are satisfied, it is called a measure
of the resource.

As concrete versions of the above general framework of resource theory, we substitute $R$ by $P, E, C,$ respectively, and  refer the corresponding cases to be resource theory of partial coherence, entanglement, and coherence, respectively. A function $C$ defined on $\cD(\cH)$ is called a {\em coherence measure} if it satisfies (C1)-(C3), and similarly we call {\em partial coherence measure} $C_a (C_b)$ and {\em entanglement measure} $E$. See Table \ref{tab-resources}.

\begin{table*}[]
\begin{tabular}{|l|l|l|l|l|}
\hline
Resource          & Free states               & Free operations               & Conditions & Measure      \\ \hline \hline
Coherence         & Incoherent states         & Incoherent operations         & C1-C3              & $C$            \\ \hline
Partial coherence & Partial incoherent states & Partial incoherent operations & P1-P3              & $C_a  ~(C_b)$ \\ \hline
Entanglement      & Separable states          & LOCC                          & E1-E3              & $E$            \\ \hline
\end{tabular}
\caption{Free states, free operations and conditions for various resource theories. Two essential components of any resource theory are free states, which do not contain any resource, and free operations, which map free states to free states. For a particular resource theory, a mathematical function of density matrix, say $R(\rho)$, satisfying the conditions (R1) faithfulness, (R2a) monotonicity under free operation, (R2b) monotonicity under selective measurement, and (R3) convexity, is called a measure of resource.When $R$ is substituted by $C$,$C_a$,$E$, the resource is refereed to as coherence, partial coherence, entanglement, respectively, }
\label{tab-resources}
\end{table*}

\subsection{Symmetric concave functions}

Before discussing our main results, we connect coherence measures with functions of row vectors.
Let $$\Omega = \Big \{p = (p_1, p_2, \cdots, p_d) \Big | \sum_{i=1}^d p_i = 1,  \  p_i \geq0 \Big \}$$
be the probability simplex of a $d$-dimensional system.
A function $f$ on $\Omega$ is called a symmetric concave function if it satisfies the following conditions:

(F1) {\sl Faithfulness}: $f(1,0,\cdots, 0)=0$.

(F2) {\sl Invariance} under any permutation transformation $P_\pi$: $f(P_\pi(p)) = f(p)$ for every $p\in \Omega$;

(F3) {\sl Concavity}: $f(rp+(1-r)q)\geq rf(p)+(1-r)f(q)$ for any $r\in[0,1]$ and $p,q\in \Omega$.

We denote by $\cF_{\rm sc}$ the set of all symmetric concave functions on the probability simplex $\Omega.$

In Ref. \cite{Du}, it was shown that coherence measures can be established through symmetric concave functions, and that all coherence measures satisfying (C1)-(C3) restrict to pure states can determine the corresponding symmetric concave functions. We focus here on the second result, that is, for any coherence measure $C$, there is a symmetric concave function $f$ satisfying (F1)-(F3) such that $C(\ket{\psi}) = f(p)$, where $p = (|p_1|^2, |p_2|^2, \cdots, |p_d|^2) $ is the {\it coherence vector} in the reference basis $\{\ket{i}\}$, i.e., $\ket{\psi} = \sum_i p_i\ket{i}$.


Any symmetric concave function is Schur-concave \cite{note1} in the sense that \cite{Du, Bhatia}
\be\label{property1}
f(p)\geq f(q) \qquad \textmd{if} \quad \ p\prec q.
\ee
Here $``\prec"$ is the majorization relation between vectors \cite{Watrous}.
%
%
Recall that  for two $n$-dimensional real vectors $x=(x_1,\cdots ,x_n) $ and $y=(y_1,\cdots ,y_n) $, $x\prec y$ means that  $\sum_{i=1}^n x_i=\sum_{i=1}^ny_i$ and
 \begin{align*}
 \sum^k_{i=1} x^{\downarrow}_i \le \sum^k_{i=1}y^{\downarrow}_i, \qquad k=1,\cdots ,n-1,
 \end{align*}
where $x^{\downarrow} \equiv (x^{\downarrow}_1,\cdots ,x^{\downarrow}_n)$ and $y^{\downarrow} \equiv (y^{\downarrow}_1,\cdots ,y^{\downarrow}_n)$ respectively are the non-increasing rearrangements of $x$ and $y$.
In this case, $x$ is said to be majorized by $y.$

\section{Transformation and measures of partial coherence}

\subsection{Transformation of partial coherence}

In Ref. \cite{Nielsen}, Nielsen showed that a  bipartite entangled pure state  can be transformed to another one if and only if the vectors of their Schmidt numbers have a majorization relationship. Along this spirit, Du {\sl et al.} constructed a similar result in coherence theory: A state can be transformed to another state  if and only if the coherence vector of the former is majorized by that of the latter \cite{Du2}. Here we establish a corresponding result for partial coherence, which extends the theorem of Du {\sl et al.} \cite{Du2}.

Consider a bipartite system shared by parties $a$ and $b$, and  fix an orthonormal basis $\{\ket{i}: i=1, 2 ,\cdots ,d_a\}$ of party $a$. The partial coherence vector of a bipartite pure state $$\ket{\psi}=\sum_{i,j} \psi_{i,j}\ket{i}_a\ket{j}_b$$
 is defined as
\begin{align}\label{partial-coherence-vector}
\psi_a=\Big (\sum_j|\psi_{1,j}|^2, \sum_j|\psi_{2,j}|^2, \cdots, \sum_j|\psi_{d_a,j}|^2\Big ) .
\end{align}
Clearly, the above pure state can be rewritten as
\begin{align*}
\ket{\psi} = \sum_{i} \psi_{i}\ket{i}_a\ket{\psi_i}_b
\end{align*}
with
$$\psi_{i}=\sqrt{\sum_j|\psi_{i,j}|^2}, \quad \ket{\psi_i}_b = \sum_j \frac{\psi_{i,j}}{\psi_{i}}\ket{j}_b.
$$
The partial coherence vector can then be recast as
\begin{align}\label{simple-pcv}
\psi_a=(\psi_1^2, \psi_2^2, \cdots, \psi_{d_a}^2).
\end{align}
The fact $\psi_{i}^2=\bra{i}\textmd{tr}_b(\ket{\psi}\bra{\psi})\ket{i}$ implies that the partial coherence vector does not actually depend on the choice of the basis of party $b$.

{\bf Theorem 1}. A bipartite pure states $\ket{\psi}$ can be transformed to another bipartite pure state $\ket{\varphi}$ by partial incoherent operations  (PIO) if and only if $\psi_a$ is majorized by $\varphi_a$. Symbolically,
\begin{equation}
\label{marjority-relation-of-partial-coherence}
\ket{\psi}{\tiny{\overrightarrow{\quad {\rm PIO}\quad }}}\ket{\varphi} \Longleftrightarrow  \psi_a \prec  \varphi_a.
\end{equation}

For proof, see Appendix {\bf A}.

An immediate corollary of the above theorem  is the following result:  Let $\rho$ be any state (pure or mixed) of $\mathcal{H}_{ab}$, and
$$\ket{\psi_{\rm max}}=\frac{1}{\sqrt{d_a}}\sum_i\ket{i}_a\ket{\psi_i}_b.$$
where $\ket{\psi_i}$ are arbitrary quantum states of party $b$ (not necessarily orthogonal to each other).
 Then there is a partial incoherent operation transforming $\ket{\psi_{\rm max}}$ to $\rho.$

This motivates the name maximal partial coherent state for $\ket{\psi_{\rm max}}.$  Consider a bipartite state with the pure-state decomposition $$\rho=\sum_iq_i\ket{y_i}\bra{y_i}.$$
Since the partial coherence vector of any maximal partial coherent state is $\psi_a=(\frac{1}{d_a},\cdots ,\frac{1}{d_a}) ,$ which is majorized by the partial coherence vector of each $\ket{y_i}$, Theorem 1 implies that for each $i$, there exists a partial incoherent operation ${\cal E}_i$ which transforms $\ket{\psi_{\rm max}}$ to  $\ket{y_i}$. Therefore, the partial incoherent operation ${\cal E}(\cdot)=\sum_iq_i{\cal E}_i(\cdot)$ maps $\ket{\psi_{\rm max}}$ to the target state $\rho$. In this sense, any bipartite pure state with the partial coherence vector $(\frac{1}{d_a},\cdots ,\frac{1}{d_a}) $ is a maximal partial coherent state, because it can be used to prepare any state in a way that does not consume additional coherence resource.

From Theorem 1, we further obtain that each maximal partial coherent state has the maximal value for each partial coherence measure $C_a $, because the property (P2a) with $P$ replaced by $C$ implies $C_a (\ket{\psi_{\rm max}})\ge C_a (\rho).$

As is well known, majorization is  only a partial order relationship. Supplementary to Theorem 1,  there can exist two pure states $\ket{\psi}$ and $\ket{\varphi}$ with neither $\ket{\psi}\tiny{\overrightarrow{\quad {\rm PIO} \quad }}\ket{\varphi}$ nor $\ket{\varphi}\tiny{\overrightarrow{\quad {\rm PIO} \quad }}\ket{\psi}$. For instance, let $d_a=4$ and
\begin{align*}
\ket{\psi}&=\sqrt{0.5}\ket{1}_a\ket{\psi_1}_b+\sqrt{0.26}\ket{2}_a\ket{\psi_2}_b+\sqrt{0.24}\ket{3}_a\ket{\psi_3}_b,\\
\ket{\varphi}&=\sqrt{0.4}\ket{1}_a\ket{\varphi_1}_b+\sqrt{0.4}\ket{2}_a\ket{\varphi_2}_b+\sqrt{0.15}\ket{3}_a\ket{\varphi_3}_b\\
&+\sqrt{0.05}\ket{4}_a\ket{\varphi_4}_b
\end{align*}
where $\ket{\psi_i}_b$ and $\ket{\varphi_j}_b$ are arbitrary quantum states of system $b$.
Then it is easy to check that there are no {\rm PIO} connecting them.

For entanglement and coherence theory, a remarkable feature is catalysis \cite{Jonathan1999}. This allows a conversion between two initially non-convertible entangled (coherent) states by a borrowed entangled (coherent) state, which is recovered at the end of the process. This borrowed entangled (coherent) state is called an entangled (a coherent) catalyst. In partial coherence theory, we can define the partial coherent catalyst and it is worth noting that the local coherent state of party $a$ can be also used to catalyze the conversion of quantum states. For example, the coherent state $\ket{\phi}=\sqrt{0.6}\ket{1}_{a^{\prime}}+\sqrt{0.4}\ket{2}_{a^{\prime}}$ is a  catalysis of the above two states, i.e., it holds that $\ket{\varphi}\ket{\phi}\tiny{\overrightarrow{\quad {\rm PIO} \quad }}\ket{\psi}\ket{\phi}$, and the corresponding partial coherent state
$$\ket{\tilde{\phi}}=\sqrt{0.6}\ket{1}_{a^{\prime}}\ket{\phi_1}_{ab}+\sqrt{0.4}\ket{2}_{a^{\prime}}\ket{\phi_2}_{ab}$$
is also a partial coherent catalyst.

We further have the following consequence of Theorem 1: For bipartite pure states $\ket{\psi}=\sum_i\psi_i\ket{i}_a\ket{\psi_i}_b$ and $\ket{\varphi}=\sum_i\varphi_i\ket{i}_a\ket{\varphi_i}_b$, $\ket{\psi}\tiny{\overrightarrow{\quad {\rm PIO} \quad }}\ket{\varphi}$ if and only if $\ket{\psi^{\prime}}\tiny{\overrightarrow{\quad {\rm IO} \quad }}\ket{\varphi^{\prime}},$ where $\ket{\psi^{\prime}}=\sum_i\psi_i\ket{i}$ and $\ket{\varphi^{\prime}}=\sum_i\varphi_i\ket{i}$.
The catalytic partial coherence transformation theory can be derived directly from the coherence catalyst \cite{Bu2016}.

\subsection{Measures of partial coherence}

Symmetric concave functions play an important role in  quantification of quantum correlations. In particular, quantification of both  entanglement and coherence  has a close relationship with symmetric concave functions \cite{Qi,Ma}. Here we extend the relationship to partial coherence theory \cite{Luo,Kim}.

First, we note that a pure state $\ket{\psi}$ is a partial incoherent state if and only if $\psi_a =P_\pi(1, 0, \cdots, 0)^T $ for a permutation matrix $P_\pi$ corresponding to a permutation $\pi$. This implies that the restriction of any partial coherence measure to pure states can be derived from a function of the partial coherence vectors.
We extend the correspondence between coherence measures and symmetric concave functions \cite{Du} to measures of partial coherence  as follows.

{\bf Theorem 2}. For any symmetric concave function $f\in \cF_{\rm sc}$ and bipartite pure state $\ket{\psi}$, let
$${C_a}(\ket{\psi})= f(\psi_{a})$$
 and extend it to all density operators of $\cH_{ab}$ as
\be\label{2}
{C_a}(\rho) = \min_{p_i, \ket{\psi_i}}\sum_ip_i{C_a}(\ket{\psi_i}),
\ee
where the minimization is  over all  pure-state ensembles of $\rho = \sum_i p_i\out{\psi_i}{\psi_i}$. Then  ${C_a}$ is a measure of partial coherence (relative to party $a$). Conversely, the restriction of any  measure ${C_a}$  of partial coherence to pure states can be  derived from a function $f\in \cF_{\rm sc}$ such that
\be\label{1}
{C_a}(\ket{\psi}) = f(\psi_a)
\ee

For proof, see Appendix {\bf B}.

Actually, each measure of partial coherence  ${C_a}$ has the same correspondence structure as general coherence measure, except that it is localized. This allows us to prove Theorem 2 using a  method similar to the proof in Ref. \cite{Du}. This extended framework applies equally to the measures of  partial coherence  on party $b$, i.e., ${C_b}(\ket{\psi}) = f(\psi_b)$ where $\psi_b =(\sum_i|\psi_{i,1}|^2,\sum_i|\psi_{i,2}|^2, \cdots, \sum_i|\psi_{i,d_b}|^2) $ and $f\in \cF_{\rm sc}$.

\section{Partial coherence vs. entanglement}


In this section, we establish some relationships between partial coherence and entanglement.

Let $\{\ket{i}_a\},\{\ket{j}_b\}$ be fixed bases of the Hilbert spaces $\cH_{a},\cH_{b}$ of parties $a, b$, respectively, and define the incoherent states as the diagonal density matrices in the basis $\{\ket{i}_a\ket{j}_b\}$ of the bipartite space $\cH_{ab}$. For any bipartite pure state $$\ket{\psi}=\sum_{i=1}^{d_a}\sum_{j=1}^{d_b} \psi_{i,j}\ket{i}_a\ket{j}_b,$$
the corresponding Schmidt decomposition is 
\bea
\ket{\psi}=\sum_{n=1}^d \sqrt{p_n}\ket{\nu_n}_a\ket{\nu_n}_b, \qquad  d = \min \{d_a,d_b\}.
\eea
Thus, we have the Schmidt vector
$$p = (p_1, p_2, \cdots, p_d, 0, \cdots, 0) $$
and the \emph{coherence vector}
$$\psi = (|\psi_{1,1}|^2, \cdots, |\psi_{1,d}|^2, |\psi_{2,1}|^2, \cdots, |\psi_{d,d}|^2) $$
    of the state $\ket{\psi}.$ Clearly, the coherence vector is majorized by the partial coherence vector, i.e., $\psi\prec \psi_a$. Moreover, since we have  $\psi_a \prec p$ from  Theorem 11.8 in Ref.  \cite{Watrous} (see also Appendix {\bf C}),  for any (partial) coherence measure $(C_a)~C$ derived from the same symmetric concave function $f\in\cF_{\rm sc}$, Eq.~(\ref{property1}) implies that
\be\label{ineq1}
{C}(\ket{\psi}) = f(\psi) \geq {C_a}(\ket{\psi}) = f(\psi_a) \geq f(p).\ \
\ee

Since  entanglement of any pure state can be quantified as a function of the Schmidt vector, inequality (\ref{ineq1}) suggests a clue to relate coherence with entanglement. We here define a function with a minimum of partial coherence over all local unitary operations,  and prove that the convex-roof extension is an entanglement measure.

Let $U_a, U_b$ be unitary operators on spaces $\cH_{a},\cH_{b}$, respectively, and ${C_a}$ be a measure of partial coherence  on the composite system $\cH_{ab}$. We define
\begin{eqnarray}\label{eq1}
{E_{C_a}}(\ket{\psi})= \min_{U_{a}} {C_{a}}(U_{a}\ot{{\bf 1}_b}\ket{\psi}),
\end{eqnarray}
where the minimum is over all unitary operators $U_{a}$ on space $\cH_{a}$. Similarly, we can define ${E_{C_b}}$ as well. Moreover, denote $U_{ab} = U_a\ot U_b$, and let $C$ be a coherence measure, define
\be\label{eq1}
{E_C}(\ket{\psi})=\min_{U_{ab}} C(U_{ab}\ket{\psi})
\ee
where the minimum is over all local unitary operators $U_a\otimes U_b$.

Based on Theorem 2 and the correspondence between measures of partial coherence  and symmetric concave functions revealed in Refs. \cite{Du,Zhu}, we have the following result.

{\bf Theorem 3}.  Let $p$ be the Schmidt vector of $\ket{\psi}$, i.e., $\ket{\psi}= \sum_i \sqrt{p_i}\ket{\nu_i}_a\ket{\nu_i}_b$ and $p = (p_1, p_2, \cdots, p_d, 0, \cdots, 0) $.
Then, for any partial coherence measure ${C_a}(\ket{\psi}) = f(\psi_a)$ with partial coherence vector $\psi_a$,  we have
\bea
{E_{C_a}}(\ket{\psi})={E_{C_b}}(\ket{\psi})= f(p)= {E_C}(\ket{\psi}).
\eea

For proof, see Appendix {\bf D}.

From Theorem 3, we know that the domain of the function $f$ associated with ${E_{C_a}}$ for all pure states corresponds one-to-one to the $d$-dimensional probability vectors, which allows us to reduce the Schmidt vector $p$ to the $d$-dimensional spectral vector without loss of generality, that is, $p = (p_1, p_2, \cdots, p_d, 0, \cdots, 0)  \rightarrow (p_1, p_2, \cdots, p_d) $.

For any bipartite quantum state $\rho \in \mathcal{D}(\cH_{a}\ot\cH_{b})$, let
\bea
{E_{C_a}}(\rho)=\min_{\lambda_i,\ket{\psi_i}}\sum_i \lambda_i{E_{C_a}}(\ket{\psi_i})
\eea
and
\bea
{E_C}(\rho)=\min_{\lambda_i,\ket{\psi_i}}\sum_i \lambda_i{E_C}(\ket{\psi_i})
\eea
where the minimum is over all possible decompositions $\rho = \sum_i \lambda_i \out{\psi_i}{\psi_i}$.

\begin{figure}%
\includegraphics[width=3in]{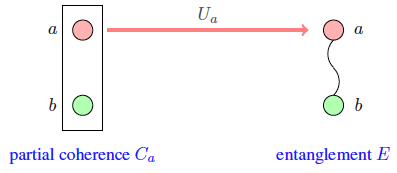}%
\caption{Entanglement of pure states can be viewed as the partial coherence that cannot be erased by local unitary operations acting on party $a$. This is also true for partial coherence $C_b$ on party $b$.}
\label{fig:fig1}
\end{figure}

Obviously, Theorem 3 implies that ${E_{C_a}}(\rho)={E_C}(\rho)$ for every quantum state $\rho\in \mathcal{D}(\cH_{a}\ot\cH_{b})$.
We here consider the correspondence between the following functions, and show that ${E_C}$ is an entanglement measure for any measure of partial coherence  ${C}$:
Let the function $g_f$ be defined as $g_f(\rho_a) = {E_C}(\ket{\psi})$ for the reduced state $\rho_a = \ptr{b}{\out{\psi}{\psi}}$ of a pure state $\ket{\psi}$. Then we also have $g_f(\rho_a) = f(p)$ from Theorem 3, where $p$ is the Schmidt vector of $\ket{\psi}$ ($p$ is also the spectral vector of $\rho_a$, i.e., $p= (p_1, p_2, \cdots, p_d) $ and $\rho_a = \sum_np_n\out{\nu_n}{\nu_n}$).
We have the following result connected with  the correspondence between these functions. (See Fig. \ref{fig:fig1}.)

{\bf Theorem 4}. For any (partial) coherence measure $({C_a})~{C}$,  $({E_{C_a}})~{E_C}$ is an entanglement measure, that is, it is faithful in separable states, strong monotonous and convex.

For proof, see Appendix {\bf E}.

In general, the coherence in quantum states may increase or decrease after local unitary transformations. However, from the above theorem, we conclude that the initial coherence does not decrease to less than the degree of entanglement via local unitary transformation. This also means that it is possible to quantify entanglement as the minimal coherence after all possible unitary transformations.
In addition, we can derive entanglement measures more easily if we use measures of partial coherence. This is because our result requires only the minimum over all partial unitary operators for party $a$ or $b$, and not the minimum over all local unitary operators for the composite system  $ab$. Moreover, regardless of the choice of party $a$ or $b$, we can derive the same entanglement measures.

Now we consider connections, in the reverse direction, of entanglement and (partial) coherence.

{\bf Theorem 5}. The restriction of any entanglement measure $E$ to pure states $\ket{\psi}$ can be rewritten as Eq. (\ref{eq1}), i.e.,
\bea
\rE(\ket{\psi}) &=& \min_{U_{ab}}{C}(U_{ab}\ket{\psi})\\
&=&\min_{U_{a}}{C_{a}}(U_{a}\ot{{\bf 1}_b}\ket{\psi})\\
\big\{\textmd{or}\  &=&\min_{U_{b}}{C_{b}}(\id_{a}\ot U_{b}\ket{\psi})\big\},
\eea
where $C$ is a coherence measure and ${C_{a}}$(or ${C_{b}}$) is a measure of partial coherence  on party $a$ (or $b$).

Note that coherence measure $C$ is independent of the choice of basis for determining the diagonal matrix, as long as the basis is the tensor product of the local basis on each of the two systems that share the entanglement.

For proof, see Appendix {\bf F}.



\begin{figure}%
\includegraphics[width=3in]{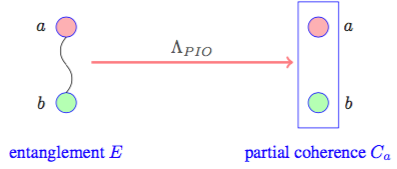}%
\caption{Partial coherence is the maximal entanglement generated by partial incoherent operations $\Lambda_{PIO}$ on party $a$. This is also true for partial coherence $C_b$ on party $b$.}
\label{fig:fig2}
\end{figure}

We have seen that entanglement can be regarded as the partial coherence that cannot be erased by local unitary operations. In a dual fashion, we now show that partial coherence is the maximal entanglement generated by partial incoherent operations (see Fig. \ref{fig:fig2}).

{\bf Theorem 6}. Let $f$ be the corresponding symmetric concave function of any entanglement measure $E$ and define $$E_f(\rho) =\min_{\lambda_i,\ket{\psi_i}}\sum_i\lambda_i E_f(\ket{\psi_i}) = \min_{\lambda_i,\ket{\psi_i}}\sum_i\lambda_if(p_i),$$
where each $p_i$ is the Schmidt vectors of $\ket{\psi_i}$. Then, for any pure state $\ket{\psi}$,
the maximal entanglement generated by partial incoherent operations is actually the induced partial coherence measure,
\bea
C_{a,f}(\ket{\psi})=f(\psi_a)=\max_{\Lambda_{PIO}} E_f\big\{\Lambda_{PIO}(\out{\psi}{\psi})\big\},
\eea
where $\psi_a$ is the partial coherence vector of $\ket{\psi}$ and the maximum is taken over all partial incoherent operations on party $a$.


For proof, see Appendix {\bf G}.

This result can be straightforwardly generalized to mixed states by convex-roof extension, i.e.,
\bea
C_{a,f}(\rho)=\min_{\lambda_i,\ket{\psi_i}}\Big[\sum_i \lambda_i\max_{\Lambda_{PIO}} E_f\big\{\Lambda_{PIO}(\out{\psi_i}{\psi_i})\big\}\Big].
\eea
The same holds true for partial coherence on party $b$.

Combining Theorems 4, 5 and 6, we establish a one-to-one correspondence between partial coherence and entanglement: entanglement is the partial coherence that cannot be erased by local unitary operations, and partial coherence is the maximal entanglement generated by partial incoherent operations.

\section{Summary}\label{8}

Both coherence and entanglement are widely used quantum resources for quantum information processing, which have been studied both extensively and intensively from the quantitative perspective in the last decades. By considering partial coherence in bipartite systems, we have investigated intrinsic connections between coherence and entanglement.
We have showed that measures of partial coherence can be constructed via arbitrary symmetric concave functions, as shown in resource theory of coherence, and have related them to entanglement measures.

We have characterized some necessary and sufficient conditions for the conversion between bipartite pure states in terms of majorization of the corresponding partial coherence vectors. This extends the corresponding results in entanglement theory. 
For pure states, while entanglement is the minimal partial coherence generated by local unitary operations, partial coherence is the maximal entanglement generated by partial incoherent operations. In addition, there is a one-to-one correspondence between the measures of partial coherence and those of entanglement.

Some interesting issues arise as the operational significance and experimental usage of the measures of partial coherence, which require further investigations.
The interplay between partial coherence and entanglement corroborates the fundamental principle of superposition, which underlies both the coherence theory and the entanglement theory. We hope the present work will be helpful in studying  entanglement in terms of partial coherence.

\begin{acknowledgments}

This project is supported by the National Natural Science Foundation of China (Grants No.12050410232, No. 12201555, No. 12031004, No. 61877054, and No. 12271471), China Postdoctoral Science Foundation (Grant No. 2021M702864) and the Fundamental Research Foundation for the Central University (Project No. K202103371).
\end{acknowledgments}

\vskip 0.3cm

\begin{center}
{\bf Appendix}
\end{center}

{\bf A. Proof of Theorem 1}

Firstly, for the sake of convenience of proof, we can express the pure states $\ket{\psi}$ and $\ket{\varphi}$ as
\be
\ket{\psi} = \sum_{i}\psi_{i}\ket{i}_a\ket{\psi_i}_b, \quad  \ket{\varphi} = \sum_{i}\varphi_{i}\ket{i}_a\ket{\varphi_i}_b
\ee
with the real numbers $\psi_{i}, \varphi_{i}$ and unit vectors
$$\ket{\psi_i}_b = \sum_j \frac{\psi_{i,j}}{\psi_{i}}\ket{j}_b, \qquad \ket{\varphi_i}_b = \sum_j \frac{\varphi_{i,j}}{\varphi_{i}}\ket{j}_b$$
 and
 $$\sum_j |\frac{\psi_{i,j}}{\psi_{i}}|^2 = \sum_j |\frac{\varphi_{i,j}}{\varphi_{i}}|^2 = 1.$$
The partial coherence vectors $\psi_a$ and $\varphi_a$ can be rewritten as $\psi_a = (\psi_{1}^2, \psi_{2}^2, \cdots, \psi_{d_a}^2) $ and $\varphi_a = (\varphi_{1}^2, \varphi_{2}^2, \cdots, \varphi_{d_a}^2) $. Here, we assume that $\psi_{1}^2 \geq \psi_{2}^2\geq \cdots\geq \psi_{d_a}^2$ and $\varphi_{1}^2 \geq \varphi_{2}^2 \geq \cdots \geq \varphi_{d_a}^2$ without loss of generality.

We here employ proof-techniques similar to those of ``only if" part in Refs. \cite{Du2,Du3}, and consider only the case when the Hilbert space of party $a$ is three-dimensional; other cases can be treated similarly. Thus suppose that $d_a=\dim \cH_{a} = 3$, $\dim \cH_{b} = d_b\geq2$ and there is a partial incoherent operation $\Phi$ that transforms $\out{\psi}{\psi}$ to $\out{\varphi}{\varphi}$. Let
\be
\Phi(\out{\psi}{\psi}) = \sum_nK_n\out{\psi}{\psi}K^\dagger_n = \out{\varphi}{\varphi}.
\ee
Hence there exist complex numbers $\alpha_n$ such that $K_n\ket{\psi} = \alpha_n\ket{\varphi}$. Let $k^{(n)}_j\ (j = 1,2,3)$ be the nonzero element at the $j$th column of $K_n$ on party $a$ (if there is no nonzero element in the $j$th column on party $a$, then $k^{(n)}_j=0$). Suppose $k^{(n)}_j$ locates the $f_n(j)$th row. Here, $f_n(j)$ is a function that maps $\{2,3\}$ to $\{1,2,3\}$ with the property that $1 \leq f_n(j) \leq j$. Let $\delta_{s,t} = 1\ (\textmd{if}\ s=t)$ or $0\ (\textmd{if}\ s\neq t)$.
Then there is a permutation $\pi_n$ such that
\begin{widetext}
\bea
K_n = P_{\pi_n}\ot \id_b\left(
                          \begin{array}{ccc}
                            k_1^{(n)}B^{(n)}_{11} &\ \delta_{1,f_n(2)}k_2^{(n)}B^{(n)}_{12} \ & \delta_{1,f_n(3)}k_3^{(n)}B^{(n)}_{13} \\
                            & &\\
                            0 & \delta_{2,f_n(2)}k_2^{(n)}B^{(n)}_{22} & \delta_{2,f_n(3)}k_3^{(n)}B^{(n)}_{23} \\
                            & &\\
                            0 & 0 & \delta_{3,f_n(3)}k_3^{(n)}B^{(n)}_{33} \\
                          \end{array}
                        \right)
\eea
where $B^{(n)}_{ij}\ (i,j = 1,2,3)$ are arbitrary matrices on party $b$.
From $\sum_n K_n^\dagger K_n = \id_{ab} (= \id_{a}\ot{{\bf 1}_b})$, we get
\begin{eqnarray}\label{16}
&&\sum_n |k_j^{(n)}|^2(B_{jj}^{(n)})^\dagger B_{jj}^{(n)} ={\bf 1}_b,  \qquad j=1,2,3,\nonumber\\
&&\sum_n \overline{k_1^{(n)}}\delta_{1,f_n(2)}k_2^{(n)}(B_{11}^{(n)})^\dagger B_{12}^{(n)} = 0,\nonumber\\
&&\sum_n \overline{k_1^{(n)}}\delta_{1,f_n(3)}k_3^{(n)}(B_{11}^{(n)})^\dagger B_{13}^{(n)} = 0,\\
&&\sum_n \overline{k_2^{(n)}}k_3^{(n)}\Big\{\delta_{1,f_n(2)}\delta_{1,f_n(3)}(B_{12}^{(n)})^\dagger B_{13}^{(n)} + \delta_{2,f_n(2)}\delta_{2,f_n(3)}(B_{22}^{(n)})^\dagger B_{23}^{(n)}\Big\} = 0.\nonumber
\end{eqnarray}
By a direct computation, we have
\bea
K_n\ket{\psi} = P_{\pi_n}\ot \id_b\left(\small{
                          \begin{array}{c}
                            k_1^{(n)}\psi_1\bra{1}B^{(n)}_{11}\ket{\psi_1}_b + \delta_{1,f_n(2)}k_2^{(n)}\psi_2\bra{1}B^{(n)}_{12}\ket{\psi_2}_b + \delta_{1,f_n(3)}k_3^{(n)}\psi_3\bra{1}B^{(n)}_{13}\ket{\psi_3}_b \\
                            \vdots\\
                            k_1^{(n)}\psi_1\bra{d_B}B^{(n)}_{11}\ket{\psi_{1}}_b + \delta_{1,f_n(2)}k_2^{(n)}\psi_2\bra{d_B}B^{(n)}_{12}\ket{\psi_2}_b + \delta_{1,f_n(3)}k_3^{(n)}\psi_3\bra{d_B}B^{(n)}_{13}\ket{\psi_3}_b \\
                            \\
                            \delta_{2,f_n(2)}k_2^{(n)}\psi_2\bra{1}B^{(n)}_{22}\ket{\psi_2}_b + \delta_{2,f_n(3)}k_3^{(n)}\psi_3\bra{1}B^{(n)}_{23}\ket{\psi_3}_b \\
                            \vdots\\
                            \delta_{2,f_n(2)}k_2^{(n)}\psi_2\bra{d_B}B^{(n)}_{22}\ket{\psi_2}_b + \delta_{2,f_n(3)}k_3^{(n)}\psi_3\bra{d_B}B^{(n)}_{23}\ket{\psi_3}_b \\
                            \\
                            \delta_{3,f_n(3)}k_3^{(n)}\psi_3\bra{1}B^{(n)}_{33}\ket{\psi_3}_b \\
                            \vdots\\
                            \delta_{3,f_n(3)}k_3^{(n)}\psi_3\bra{d_B}B^{(n)}_{33}\ket{\psi_3}_b
                          \end{array}}
                        \right)
\eea
and hence
\begin{eqnarray}\label{17}
    &&\quad k_1^{(n)}\psi_1B^{(n)}_{11}\ket{\psi_1}_b + \delta_{1,f_n(2)}k_2^{(n)}\psi_2B^{(n)}_{12}\ket{\psi_2}_b + \delta_{1,f_n(3)}k_3^{(n)}\psi_3B^{(n)}_{13}\ket{\psi_3}_b = \alpha_n\varphi_{\pi_n^{-1}(1)}\ket{\varphi_{\pi_n^{-1}(1)}}_b,\qquad\qquad\nonumber \\
    &&\quad \delta_{2,f_n(2)}k_2^{(n)}\psi_2B^{(n)}_{22}\ket{\psi_2}_b + \delta_{2,f_n(3)}k_3^{(n)}\psi_3B^{(n)}_{23}\ket{\psi_3}_b = \alpha_n\varphi_{\pi_n^{-1}(2)}\ket{\varphi_{\pi_n^{-1}(2)}}_b, \nonumber\\
    &&\quad \delta_{3,f_n(3)}k_3^{(n)}\psi_3B^{(n)}_{33}\ket{\psi_3}_b = \alpha_n\varphi_{\pi_n^{-1}(3)}\ket{\varphi_{\pi_n^{-1}(3)}}_b,
\end{eqnarray}
where, for $s=1,2,3$,
\bea
\varphi_{\pi_n^{-1}(s)}\ket{\varphi_{\pi_n^{-1}(s)}}_b = \delta_{\pi_n^{-1}(s),1}\varphi_{1}\ket{\varphi_{1}}_b + \delta_{\pi_n^{-1}(s),2}\varphi_{2}\ket{\varphi_{2}}_b +\delta_{\pi_n^{-1}(s),3}\varphi_{3}\ket{\varphi_{3}}_b.
\eea
After summing up the inner products of equations in (\ref{17}) for all $n$ and organizing them using (\ref{16}), we have the following three equations:
\bea
&i)&\quad\psi_1^2 + \sum_n\delta_{1,f_n(2)}|k_2^{(n)}|^2\psi_2^2\bra{\psi_2}(B^{(n)}_{12})^\dagger B^{(n)}_{12}\ket{\psi_2}_b + \sum_n\delta_{1,f_n(3)}|k_3^{(n)}|^2\psi_3^2\bra{\psi_3}(B^{(n)}_{13})^\dagger B^{(n)}_{13}\ket{\psi_3}_b\\
&&+\sum_n\delta_{1,f_n(2)}\delta_{1,f_n(3)}\overline{k_2^{(n)}}k_3^{(n)}\psi_2\psi_3\bra{\psi_2}(B^{(n)}_{12})^\dagger B^{(n)}_{13}\ket{\psi_3}_b + \sum_n\delta_{1,f_n(2)}\delta_{1,f_n(3)}\overline{k_3^{(n)}}k_2^{(n)}\psi_3\psi_2\bra{\psi_3}(B^{(n)}_{13})^\dagger B^{(n)}_{12}\ket{\psi_2}_b\\
&&= \sum_n|\alpha_n|^2\varphi_{\pi_n^{-1}(1)}^2 \coloneqq \mathbf{I};\\
&ii)&\quad\sum_n\delta_{2,f_n(2)}|k_2^{(n)}|^2\psi_2^2\bra{\psi_2}(B^{(n)}_{22})^\dagger B^{(n)}_{22}\ket{\psi_2}_b + \sum_n\delta_{2,f_n(3)}|k_3^{(n)}|^2\psi_3^2\bra{\psi_3}(B^{(n)}_{23})^\dagger B^{(n)}_{23}\ket{\psi_3}_b\\
&&+\sum_n\delta_{2,f_n(2)}\delta_{2,f_n(3)}\overline{k_2^{(n)}}k_3^{(n)}\psi_2\psi_3\bra{\psi_2}(B^{(n)}_{22})^\dagger B^{(n)}_{23}\ket{\psi_3}_b + \sum_n\delta_{2,f_n(2)}\delta_{2,f_n(3)}\overline{k_3^{(n)}}k_2^{(n)}\psi_3\psi_2\bra{\psi_3}(B^{(n)}_{23})^\dagger B^{(n)}_{22}\ket{\psi_2}_b\\
 &&= \sum_n|\alpha_n|^2\varphi_{\pi_n^{-1}(2)}^2 \coloneqq \mathbf{II};\\
&iii)&\quad\sum_n\delta_{3,f_n(3)}|k_3^{(n)}|^2\psi_3^2\bra{\psi_3}(B^{(n)}_{33})^\dagger B^{(n)}_{33}\ket{\psi_3}_b = \sum_n|\alpha_n|^2\varphi_{\pi_n^{-1}(3)}^2 \coloneqq \mathbf{III}.
\eea
Note that, for $s=1,2,3$,
\bea
\sum_n|\alpha_n|^2\varphi_{\pi_n^{-1}(s)}^2 &=& \sum_{n}|\alpha_n|^2\big(\delta_{\pi_n^{-1}(s),1}\varphi_{1}^2 + \delta_{\pi_n^{-1}(s),2}\varphi_{2}^2 + \delta_{\pi_n^{-1}(s),3}\varphi_{3}^2\big)\\
 &=& \sum_{n,\pi_n^{-1}(s)=1}|\alpha_n|^2\varphi_{1}^2 + \sum_{n,\pi_n^{-1}(s)=2}|\alpha_n|^2\varphi_{2}^2 +\sum_{n,\pi_n^{-1}(s)=3}|\alpha_n|^2\varphi_{3}^2.
\eea
\end{widetext}
Let $d_{ij} = \sum_{n,\pi_n^{-1}(i)=j}|\alpha_n|^2, 1\leq i,j \leq 3.$ Then the matrix $D = (d_{ij})$ is a double stochastic matrix since $\sum_{n}|\alpha_n|^2 = 1$. And because
\bea
D(\varphi_1^2, \varphi_2^2, \varphi_3^2) = (\mathbf{I}, \mathbf{II}, \mathbf{III}) ,
\eea
we can get
\begin{eqnarray}\label{22}
(\mathbf{I}, \mathbf{II}, \mathbf{III})  \prec (\varphi_1^2, \varphi_2^2, \varphi_3^2) .
\end{eqnarray}
On the other hand, we have $\psi_1^2 \leq \mathbf{I}$ from the following equation
\bea
\mathbf{I}-\psi_1^2 = \sum_n \iinner{\omega^{(n)}}{\omega^{(n)}}_b,
\eea
where
$$\ket{\omega^{(n)}}_b = \delta_{1,f_n(2)}k_2^{(n)}\psi_2B^{(n)}_{12}\ket{\psi_2}_b + \delta_{1,f_n(3)}k_2^{(n)}\psi_3B^{(n)}_{13}\ket{\psi_3}_b.$$
And, by the first equation of (\ref{16}), we have \ $\mathbf{III} \leq \psi_3^2$. This implies that
\bea
\psi_1^2+\psi_2^2 \leq \mathbf{I}+\mathbf{II}.
\eea
From the definition of majorization, we can obtain
\begin{eqnarray}\label{18}
(\psi_1^2, \psi_2^2, \psi_3^2)  \prec (\mathbf{I}, \mathbf{II}, \mathbf{III}) .
\end{eqnarray}
By combining (\ref{22}) and (\ref{18}), we have
\bea
(\psi_1^2, \psi_2^2, \psi_3^2)  \prec (\varphi_1^2, \varphi_2^2, \varphi_3^2) .
\eea

Next we prove the ``sufficient'' part. Assume that $\psi_a\prec\varphi_a$. Firstly, for each $\ket{\psi_i}_b$, there exists a unitary operator $U_{i,b}$ satisfying $U_{i,b}\ket{\psi_i}_b=\ket{\varphi}_b$. Then the operator $U=\sum_i\ket{i}_a\bra{i}\otimes U_{i,b}$ is unitary and $\Phi_{U}(\cdot)=U\cdot U^{\dagger}$ is a partial incoherent operation on party $a$  such that
\begin{align*}
\Phi_{U}(\ket{\psi}\bra{\psi})=\Big (\sum_i\psi_i\ket{i}_a\ket{\varphi}_b \Big ) \Big (\sum_j\psi_j\ket{j}_a\ket{\varphi}_b\Big )^{\dagger}.
\end{align*}

Secondly, denote $\ket{\varphi^{\prime}}=\sum_i\varphi_i\ket{i}_a\ket{\varphi}_b$ and for each $i$, define unitary operator $U^{\prime}_{i,b}$ satisfying $U^{\prime}_{i,b}\ket{\varphi}_b=\ket{\varphi_i}_b$. Then the operator $U^{\prime}=\sum_i\ket{i}_a\bra{i}\otimes U^{\prime}_{i,b}$ is unitary and $\Phi_{U'}(\cdot)=U^{\prime}\cdot U^{\prime\dagger}$ is a partial incoherent operation on party $a$   such that
\begin{align*}
\Phi_{U'}(\ket{\varphi^{\prime}}  \bra{\varphi^{\prime}})= \ket{\varphi}\bra{\varphi}.
\end{align*}

At last, let us consider two pure states $\ket{\psi}_a=\sum_i\psi_i\ket{i}_a$ and $\ket{\varphi}_a=\sum_i\varphi_i\ket{i}_a$. Then the condition $\psi_a\prec\varphi_a$ implies that there exists an incoherent operation $\Lambda_a$ such that \cite{Du2,Zhu}
\begin{align*}
\Lambda_a(\ket{\psi}_a\bra{\psi})=\ket{\varphi}_a\bra{\varphi}.
\end{align*}
Therefore, $\Lambda_a\otimes I_b$ and $\Phi_{U^{\prime}}\circ(\Lambda_a\otimes I_b)\circ\Phi_{U}$ are both partial incoherent operations, and it holds that
\begin{align*}
&\Phi_{U^{\prime}}\circ(\Lambda_a\otimes I_b)\circ\Phi_{U}(\ket{\psi}\bra{\psi})\\
=&\Phi_{U^{\prime}}\circ(\Lambda_a\otimes I_b)\Big (\sum_i\psi_i\ket{i}_a\ket{\varphi}_b\sum_j\psi_j\bra{j}_a\bra{\varphi}_b \Big )\\
=&\Phi_{U^{\prime}}\Big (\sum_i\varphi_i\ket{i}_a\ket{\varphi}_b\sum_j\varphi_j\bra{j}_a\bra{\varphi}_b \Big )=\ket{\varphi}\bra{\varphi}.
\end{align*}

\vskip 0.2cm

{\bf B. Proof of Theorem 2}

 For any $\rho\in \cI_a$, i.e., $\rho = \sum_i p_i\ket{i}_a\bra{i}\ot\rho_{i}$, where $\rho_i = \sum_j p^{(i)}_{j}\ket{\phi^{(i)}_j}_b\bra{\phi^{(i)}_j}$, from the definition of ${C_a}$ and (F1), it follows that ${C_a}(\rho)\leq \sum_{i,j}p_ip^{(i)}_{j} C_a (\ket{i}_a\ket{\phi^{(i)}_j}_b) = 0$. On the contrary, ${C_{a}}(\rho)=0$ implies that  $\rho=\sum_iq_i\ket{\psi_i}\bra{\psi_i}$ is a decomposition of $\rho$ with ${C_{a}}(\ket{\psi_i})=0$, which means that $\ket{\psi_i}=\ket{P_\pi(i)}_a\sum_j\sqrt{q^{(P_\pi(i))}_j}\ket{\psi^{(P_\pi(i))}}_j$. It is easy to check that $\rho$ is a partial incoherent state. 





For the proof of (P2b), there is a partial incoherent operation $\Lambda_{ab}$ with Kraus operators $K_n$.
We assume firstly that $\rho$ is a pure state $\out{\psi}{\psi}$ and denote $p_n = \|K_n\ket{\psi}\|^2$, $\rho_n = \frac{1}{p_n}K_n\out{\psi}{\psi}K_n^\dagger$ and $P_{\pi_n}(\frac{|\phi^{(n)}_1|^2}{p_n}, \frac{|\phi^{(n)}_2|^2}{p_n}, \cdots, \frac{|\phi^{(n)}_{d_A}|^2}{p_n}) $ are partial coherence vectors of $\frac{K_n\ket{\psi}}{\sqrt{p_n}}$ for all $n$. By the same method as the proof of Theorem 1, we get
\be
(\psi_1^2, \cdots, \psi_{d_a}^2)  \prec \Big (\sum_n|\phi^{(n)}_1|^2, \cdots, \sum_n|\phi^{(n)}_{d_a}|^2 \Big ).
\ee
From (F2) and (F3), it follows that
\begin{align}\label{9}
& \sum_n p_n{C_{a}}(\rho_n)\nonumber \\
&= \sum_n p_n f\Big (P_{\pi_n}\Big (\frac{|\phi^{(n)}_1|^2}{p_n}, \frac{|\phi^{(n)}_2|^2}{p_n}, \cdots, \frac{|\phi^{(n)}_{d_a}|^2}{p_n}\Big ) \Big )\nonumber \\
&= \sum_n p_n f\Big (\Big (\frac{|\phi^{(n)}_1|^2}{p_n}, \frac{|\phi^{(n)}_2|^2}{p_n}, \cdots, \frac{|\phi^{(n)}_{d_a}|^2}{p_n}\Big ) \Big )\nonumber\\
&\leq f((\sum_n|\phi^{(n)}_1|^2, \sum_n|\phi^{(n)}_2|^2, \cdots, \sum_n|\phi^{(n)}_{d_a}|^2) )\\
&\leq f((\psi_1^2, \psi_2^2, \cdots, \psi_{d_a}^2) ) = {C_{a}}(\rho).\nonumber
\end{align}

The last inequality follows from (\ref{property1}).
For mixed states also, the inequality in (\ref{9}) holds.  Actually, let $\rho=\sum_iq_i\rho_i$ be the optimal pure-state ensemble
with ${C_{a}}(\rho)=\sum_iq_i{C_{a}}(\rho_i)$, and denote $\rho_n=K_n\rho K^{\dagger}_n/p_n$ with $p_n=\textmd{tr}(K_n\rho K^{\dagger}_n)$. Then, one has
\begin{align*}
&\sum_np_n{C_{a}}(\rho_n)=\sum_np_n{C_{a}} \Big (\frac{\sum_iq_iK_n\rho_iK^{\dagger}_n}{p_n} \Big)\\
&\le \sum_np_n\sum_iq_i\frac{\textmd{tr}(K_n\rho_iK^{\dagger}_n)}{p_n}{C_{a}} \Big (\frac{K_n\rho_iK^{\dagger}_n}{\textmd{tr}(K_n\rho_iK^{\dagger}_n)} \Big )\\
& =\sum_iq_i  \sum _n\textmd{tr} \Big (K_n\rho_iK^{\dagger}_n \Big ) C_a \Big (\frac{K_n\rho_iK^{\dagger}_n}{\textmd{tr}(K_n\rho_iK^{\dagger}_n)} \Big )\\
&\le \sum_iq_i{C_{a}}(\rho_i)={C_{a}}(\rho),
\end{align*}
where the inequalities follow from the definition of ${C_{a}}$ and Ineq.~(\ref{9}), respectively.

Finally, the convexity (P3) of ${C_{a}}$ is obtained directly from the definition of minimization of ${C_{a}}$. In fact, let $\rho=\sum_ip_i\sigma_i$ be any ensemble of $\rho$ and $\sigma_i=\sum_jq_{ij}\sigma_{ij}$
be the optimal pure-state ensemble of $\sigma_i$, i.e., ${C_{a}}(\sigma_i)=\sum_iq_{ij}{C_{a}}(\sigma_{ij})$. Then
\begin{align*}
&{C_{a}} \Big (\sum_ip_i\sigma_i \Big )={C_{a}} \Big (\sum_{i,j}p_iq_{ij}\sigma_{ij} \Big )\\
&\le \sum_ip_i\sum_jq_{ij}{C_{a}}(\sigma_{ij})=\sum_{i}p_i{C_{a}}(\sigma_i),
\end{align*}
where the inequality follows from the definition of ${C_{a}}$.

Now we prove the second part of the theorem. A bipartite pure state $\ket{\psi}=\sum_{i,j}\psi_{ij}\ket{i}_a\ket{j}_b$ can also be rewritten as
\begin{align*}
\ket{\psi}=\sum_{i}\psi_{i}\ket{i}_a\ket{\psi_i}_b,
\end{align*}
where $\ket{\psi_i}_b=\sum_j\frac{\psi_{ij}}{\psi_i}\ket{j}_b$ and $\psi_i=\sqrt{\sum_j|\psi_{ij}|^2}$. Since partial coherence is invariant under local unitary operations on party $b$, then for each measure of partial coherence  $C_a$, there exists a function $f$ such that the partial coherence of any pure state is a functional value of the corresponding partial coherence vectors, that is,
\begin{align*}
C_a(\ket{\psi}\bra{\psi})=f\Big (\sum_j|\psi_{1,j}|^2,\cdots ,\sum_j|\psi_{d_a,j}|^2\Big ).
\end{align*}




\vskip 0.2cm

{\bf C. Proof} of $\psi_a\prec p$

We use the Schur-Horn theorem in our proof.

{\em Schur-Horn Theorem (See also Theorem 13.7 in Ref. \cite{Watrous})}. Let $\mathcal{X}$ be a complex Euclidean space, $A$ be a Hermitian operator on $\mathcal{X}$,
and $\{\ket{x_a}\}$ be an orthonormal basis of $\mathcal{X}$. For a vector $v$ that is defined as $v(a)=\bra{x_a}
A\ket{x_a}$, it holds that
\begin{align*}
v\prec \lambda(A),
\end{align*}
where $\lambda(A)$ is the spectral vector of $A$.

Now we prove our result. Our aim is to prove that for a bipartite pure state
\begin{align*}
\ket{\psi}=\sum_{i}^{d_a}\sum_{j}^{d_b} \psi_{i,j}\ket{i}_a\ket{j}_b=\sum_{n}^d \sqrt{p_n}\ket{\nu_n}_a\ket{\nu_n}_b,
\end{align*}
the partial coherence vector $\psi_a =(\sum_j|\psi_{1,j}|^2,\sum_j|\psi_{2,j}|^2, \cdots, \sum_j|\psi_{d_a,j}|^2)$ is majorized by the spectral vector $p = (p_1, p_2, \cdots, p_{d})$ ($d = \min\{d_a,d_b\}$), i.e., $\psi_a\prec p$.

As we can see, the reduced state of the above pure state $\ket{\psi}$ can be written as $\rho_a=\sum_np_n\ket{\nu_n}_a\bra{\nu_n}$, where $\lambda(\rho_a) = p$ and each entry of partial coherence vector can be represented by $\sum_j|\psi_{i,j}|^2=\bra{i}\rho_a\ket{i}$. Then, the Schur-Horn theorem implies that
\begin{align*}
\psi_a\prec p.
\end{align*}

\vskip 0.2cm

{\bf D. Proof of Theorem 3}

First, for any pure state $\ket{\psi}$, since the vector $p$ is local unitary invariant, we have
\begin{align*}
{E_{C_a}}(\ket{\psi})=C_a(\ket{\tilde{\psi}})=f(\tilde{\psi}_a)\ge f(p),
\end{align*}
where $\ket{\tilde{\psi}}=(\tilde{U}_a\otimes {\bf 1}_b)\ket{\psi}$ reaches the minimal partial coherence and the inequality is the result of Theorem 11.8 of Ref. \cite{Watrous}.
In addition, suppose $U'_{a}$ is the local unitary operation that satisfies $U'_{a}\ket{\nu_i}_a=\ket{i}_a, i \in \{1,2,\cdots, d\}),$ then we have $p_i = \bra{i}_a\ptr{b}{\out{\psi'}{\psi'}}\ket{i}_a$ with $\ket{\psi'} = U'_{a}\ot{{\bf 1}_b}\ket{\psi}$, and it holds that
 \begin{align*}
 {E_{C_a}}(\ket{\psi})\le C_a(U'_{a}\otimes I_b\ket{\psi})=f(p).
 \end{align*}
In conclusion, we have ${E_{C_a}}(\ket{\psi})=f(p)$, and the equation also holds for ${E_{C_b}}$.

Secondly, for all $U_{ab}$, the Schmidt vector of pure state $U_{ab}\ket{\psi}$ remains unchanged, and $C(U_{ab}\ket{\psi})\geq f(p)$ is established for all $U_{ab}$ from inequality (\ref{ineq1}). Consequently, we have ${E_C}(\ket{\psi})\geq f(p)$.

Moreover, for the unitary operator $U'_{ab}$ acting on the Schmidt basis of the pure state $\ket{\psi} = \sum_i^d \sqrt{p_i}\ket{\nu_i}_a\ket{\nu_i}_b$ and yielding the reference basis for coherence, i.e., $(U'_{ab})\ket{\nu_i}_a\ket{\nu_j}_b=\ket{i}_a\ket{j}_b \ (i,j \in \{1,2,\cdots, d\})$, the Schmidt vector and the coherence vector for $\ket{\psi}$ after $U'_{ab}$ become the same, then one has
\begin{align}
\label{relation-between-entanglement-and-coherence}
{E_C}(\ket{\psi})= f(p).
\end{align}

\vskip 0.2cm

{\bf E. Proof of Theorem 4}

To prove Theorem 4, we need the following result obtained by Vidal \cite{Vidal3} : Given a real function $g$ on the density matrices $\rho \in \cD({\cH})$ such that it is invariant under any  unitary transform $U$, i.e.,
\begin{align}\label{eq2}
g(U\rho U^{\dagger})=g(\rho)
 \end{align}
 and concave, i.e.,
 \begin{align}\label{eq3}
 g(r\rho_1+(1-\lambda)\rho_2)\ge rg(\rho_1)+(1-r)g(\rho_2),
 \end{align}
for any $r\in[0,1]$ and any $\rho_1,\rho_2\in\mathcal{D}(\mathcal{H})$, then the function $\nu$ defined on density matrices of bipartite system $\mathcal{H}_a\otimes\mathcal{H}_b$ as
\begin{align*}
\nu(\rho)=\min_{p_i,\ket{\psi_i}}\sum_ip_i\nu(\ket{\psi_i}),
\end{align*}
where the minimum is taken over all pure state decompositions $\rho=\sum_ip_i\ket{\psi_i}\bra{\psi_i})$ and $\nu(\ket{\psi})=g(\ptr{b}{\ket{\psi}\bra{\psi}})$, is an entanglement monotone, that is, $\nu(\rho)$ satisfies (E2b) and (E3).

Since ${E_{C_a}}$ is equivalent to ${E_C}$, we just need to prove that the former is an entanglement measure. To prove our result, let $g_f(\rho_a) = {E_{C_a}}(\ket{\psi}) = f(p)$, where $\rho_a = \ptr{b}{\out{\psi}{\psi}}$. Thanks to the above result in Ref. \cite{Vidal3}, it is enough to  prove that $g_f$ satisfies Eqs.~(\ref{eq2}) and (\ref{eq3}), and ${E_{C_a}}$ is faithful on the set of separable states.

(i) To prove that Eq. (\ref{eq2}) holds, we first note that
  $$g_f(U\rho_a U^{\dagger}) = {E_{C_a}}(U\ot {\bf 1}_b \ket{\psi})$$ for any unitary operator $U$ because $\ptr{b}{U\ot {\bf 1}_b \out{\psi}{\psi}U^{\dagger}\ot {\bf 1}_b}=U\rho_a U^{\dagger}.$
  As a result, Eq. (\ref{eq2}) is obtained from  $E_{C_a}$  defined as the minimum of partial coherence over all possible local unitary operators, i.e., ${E_{C_a}}(U\ot {\bf 1}_b \ket{\psi}) = {E_{C_a}}(\ket{\psi})$.

(ii) Concavity: Let $\rho = r\rho_1+(1-r)\rho_2$ with $r\in[0,1]$ and $\rho_1, \rho_2\in\cD(\cH)$ and $\rho=\sum_np_n\out{\nu_n}{\nu_n}$, $\rho_1=\sum_iq_i\out{\upsilon_i}{\upsilon_i}$, $\rho_2=\sum_js_j\out{\tau_j}{\tau_j}$ be the spectral decompositions of $\rho, \rho_1, \rho_2$, respectively. Furthermore, denote the probability vectors $q'=(q'_1, q'_2, \cdots, q'_d) $ and $s'=(s'_1, s'_2, \cdots, s'_d) $ with $q'_n=\bra{\nu_n}\rho_1\ket{\nu_n}$ and $s'_n = \bra{\nu_n}\rho_2\ket{\nu_n}$, then Theorem 13.7 of Ref. \cite{Watrous} implies that $q'\prec q$ and  $s'\prec s$, where $q = (q_1, q_2, \cdots, q_d) $ and $s = (s_1, s_2, \cdots, s_d) $ are the spectral vectors of $\rho_1$ and $\rho_2$, respectively. As a result, it holds that $p=(p_1, p_2, \cdots, p_d)  = rq'+ (1-r)s'$ and 

\parbox{6.2cm}{
\begin{align*}
\qquad g_f(\rho)=f(p)&\geq rf(q') + (1-r)f(s') \\
  &\geq rf(q)+(1-r)f(s)\\
  &=rg_f(\rho_1)+(1-r)g_f(\rho_2),
\end{align*}}\hfill
\parbox{.3cm}{\begin{eqnarray}\label{eq:6}\end{eqnarray}}\\
  where the first and the last equalities follow from the definition and we use the concavity of $f$ in the second inequality, and
the third inequality is obtained from Eq. (\ref{property1}).
  Therefore, Eq. (\ref{eq3}) is established.

(iii) Faithfulness: ${E_{C_a}}(\rho)\geq 0$ is easily proved by the nonnegativity of $C_a$. To show that ${E_{C_a}}(\rho)=0$ if and only if $\rho$ is separable, we first discuss the case where $\rho$ is pure, i.e., $\rho = \out{\psi}{\psi}$. If ${E_{C_a}}(\ket{\psi}) = 0$, there is a $U_{a}$ that satisfies $C(U_{a}\ot{\bf 1}_b\ket{\psi}) = f((1, 0,\cdots, 0) ) = 0$ from the definition of ${E_{C_a}}$ and faithfulness of $f$, which shows that $\rho$ is separable. Conversely, if $\rho$ is separable, there is a $U_{a}$ such that the partial coherence vector of $U_{a}\ot{\bf 1}_b\ket{\psi}$ becomes $(1, 0,\cdots, 0),$ and thus ${E_{C_a}}(\rho)=0$ is established. For general quantum states, the desired result follows from  the convex-roof extended definition of ${E_{C_a}}$ and the corresponding result for pure states.

\vskip 0.2cm

{\bf F. Proof of Theorem 5}

We first recall a result in Ref. \cite{Zhu}: Let $g$ be a function on the space of density matrices and $f$ be another function on the probability vector space such that $g(\rho) = f(p)$, where $p$ is the spectral vector of $\rho$. Then $g$ is symmetric concave if and only if $f$ is symmetric concave.

Now let $E$ be an entanglement measure. From Theorem 3 of Ref. \cite{Vidal3} and the above result, for any pure state $\ket{\psi}$, there is a symmetric concave function $f$ that satisfies $E(\ket{\psi}) = f(p)$, where $p$ is the Schmidt vector of $\ket{\psi}$. Then we can derive the existence of a partial coherence measure $C_{a}$ that satisfies Eq. (\ref{eq1}).

\vskip 0.2cm

{\bf G. Proof of Theorem 6}

For any partial incoherent operation $\Lambda_{PIO}$, we denote entanglement and partial coherence of $\Lambda_{PIO}(\out{\psi}{\psi})$ by
\bea
E_f\big\{\Lambda_{PIO}(\out{\psi}{\psi})\big\} &=& \sum_iq_iE_f(\ket{\varphi_i}),\\
C_{a,f}\big\{\Lambda_{PIO}(\out{\psi}{\psi})\big\} &=& \sum_jr_jC_{a,f}(\ket{\psi_j}),
\eea
where $\{q_i,\ket{\varphi_i}\}$ and $\{r_j,\ket{\psi_j})\}$ respectively are the optimal pure state decompositions of $\Lambda_{PIO}(\out{\psi}{\psi})$ for entanglement and partial coherence. Then we obtain the following inequalities:
\begin{eqnarray}\label{23}
E_f\big\{\Lambda_{PIO}(\out{\psi}{\psi})\big\} &=& \sum_iq_iE_f(\ket{\varphi_i})
\leq \sum_jr_jE_f(\ket{\psi_j}) \nonumber\\
&\leq & \sum_jr_jC_{a,f}(\ket{\psi_j}) \nonumber\\
&=& C_{a,f}\big\{\Lambda_{PIO}(\out{\psi}{\psi})\big\} \nonumber\\
&\leq& C_{a,f}(\ket{\psi}),
\end{eqnarray}
where the second inequality follows from the minimal definition of $E_f$, the third inequality is obtained from Eq. (\ref{ineq1}) and Theorem 3, and we use the monotonicity under any partial incoherent operation of $C_a$ in the last inequality.

Let $\ket{\psi}=\sum_i\psi_i\ket{i}_a\ket{\psi_i}_b$ and the corresponding partial coherence vector be $\psi_a=\{|\psi_1|^2,...,|\psi_{d_a}|^2\}$. Then for each $\ket{\psi_i}_b$, there exists a unitary operator $U_{i,b}$ satisfying $U_{i,b}\ket{\psi_i}_b=\ket{\phi_i}_b$ with  $\{\ket{\phi_i}_b\}_i$ being an orthonormal basis of party $b$. As a result, the operator $U=\sum_i\ket{i}_a\bra{i}\otimes U_{i,b}$ is unitary and $\Phi_{PIO}(\cdot)=U (\cdot) U^{\dagger}$ is a partial incoherent operation on party $a$  such that
\begin{align*}
\Phi_{PIO}(\ket{\psi}\bra{\psi})=\Big (\sum_i\psi_i\ket{i}_a\ket{\phi_i}_b \Big ) \Big (\sum_j\psi_j\ket{j}_a\ket{\phi_j}_b\Big )^{\dagger}.
\end{align*}
Here we see that the Schmidt vector of $U\ket{\psi}$ is $\psi_a$ which establishes the following result:
\begin{eqnarray}\label{24}
E_f\big\{\Phi_{PIO}(\out{\psi}{\psi})\big\} = f(\psi_a) = C_{a,f}(\ket{\psi}).
\end{eqnarray}
Therefore, from Ineq. (\ref{23}) and Eq. (\ref{24}) we get
\bea
\max_{\Lambda_{PIO}} E_f\big\{\Lambda_{PIO}(\out{\psi}{\psi})\big\}= C_{a,f}(\ket{\psi}).
\eea


%

\end{document}